\documentclass[11pt,a4paper,english,utf8]{article}

\usepackage{newcent}
\usepackage[T1]{fontenc}
\usepackage{url}
\usepackage[colorlinks=TRUE,pagebackref=TRUE]{hyperref}
\usepackage{geometry}
\usepackage{cite}
\usepackage{graphicx}
\usepackage{amsmath}
\usepackage{amssymb}
\usepackage{amsbsy}
\usepackage{makeidx}

\newcommand{\ISI}{\emph{inter spike interval}\index{inter spike interval}}
\newcommand{\isi}{\emph{isi}\index{isi}}

\newcommand{\iid}{\emph{iid}\index{independent and identically distributed}}

\index{probability density function}
\index{probability density function}
\newcommand{\CDF}{\emph{cumulative distribution function}}\index{cumulative distribution function}
\index{cumulative distribution function}
\index{survivor function}
\newcommand{\ci}{\emph{conditional intensity function}}\index{intensity function}
\newcommand{\CI}{\emph{integrated conditional intensity function}}\index{integrated intensity function}
\index{smoothing spline}
\index{smoothing parameter}
\index{cubic splines}

\newcommand{\R}{\textsf{R}\index{R}}

\newcommand{\STAR}{\textsf{STAR}\index{STAR}}

\title{On Goodness of Fit Tests For Models of Neuronal Spike Trains
  Considered as Counting Processes}
\author{
  Christophe Pouzat and Antoine Chaffiol \\
  \\
  Laboratoire de Physiologie C\'er\'ebrale, CNRS UMR 8118\\
  UFR biom\'edicale de l'universit\'e Paris-Descartes\\
  45, rue des Saints-P\`eres\\
  75006 Paris, France
}
\date{October 27, 2008}

\makeindex

\begin{document}

\maketitle

\begin{abstract}
After an elementary derivation of the ``time transformation'', mapping
a counting process onto a homogeneous Poisson process with rate one, a
brief review of Ogata's goodness of fit tests is presented and a new
test, the ``Wiener process test'', is proposed. This test is based on
a straightforward application of Donsker's Theorem to the intervals of
time transformed counting processes. The finite sample properties of
the test are studied by Monte Carlo simulations. Performances on
simulated as well as on real data are presented. It is argued that due to its
good finite sample properties, the new test is both a simple and a useful
complement to Ogata's tests. Warnings are moreover given against the
use of a single goodness of fit test.  
\end{abstract}


\section{Introduction}
\label{sec:introduction}
Modelling neuronal action potential or ``spike'' trains as
realizations of counting processes / point processes has already a rather long history
~\cite{PerkelEtAl_1967,PerkelEtAl_1967b}. When this 
formalism is used to analyse actual data 
two related problems have to be addressed:
\begin{itemize}
\item The conditional intensity of the train must be estimated.
\item Goodness of fit tests have to be developed.
\end{itemize}
 Attempts at modelling directly the conditional intensity 
of the train have appeared in the mid
80s~\cite{JohnsonSwami_1983,Brillinger_1988b,ChornoboyEtAl_1988} and,
following a 15 years long ``eclipse'', the subject has recently exhibited a
renewed
popularity~\cite{KassVentura_2001,TruccoloEtAl_2005,OkatanEtAl_2005,TruccoloDonoghue_2007}. 
The introduction of theoretically motivated goodness of fit tests in
the spike train analysis field had to
wait until the beginning of this century with a paper by~\cite[Brown
et al, 2002]{BrownEtAl_2002}. This
latter work introduced in neuroscience one of the tests proposed
by~\cite[Ogata, 1988]{Ogata_1988}. All these tests are based on the fundamental
``time transformation''~\cite{Ogata_1988} / ``time
rescaling''~\cite{BrownEtAl_2002} result stating that is the
conditional intensity model is correct, then upon mapping the spike times
onto their integrated conditional intensity a homogeneous Poisson
process with rate 1 is obtained. The tests are then all looking for
deviations from homogeneous Poisson process properties. 

We will not address the estimation of the conditional intensity in
this paper in order to keep its length within bounds. Our own take at
this problem will be presented elsewhere (Pouzat, Chaffiol and Gu,
in preparation). Simple examples of conditional intensities will
be used for simulations and applied to actual data. We will focus
instead on a new goodness of fit test based on a straightforward
application of Donsker's Theorem~\cite{Billingsley_1999}. We will show
that the test has good finite sample properties and suggest that it is
a useful complement to Ogata's tests. We will moreover argue based on
real data examples in favor a systematic use of \emph{several}~tests.

The paper is divided as follows. In
Sec.~\ref{sec:heuristicTimeTransformation}, an elementary
justification of the time transformation is presented. This proof is
obtained as a limit of a discrete problem arising from a uniform
binning of the time axis. This discretized problem is moreover the
canonical setting appearing in (nearly) every presently proposed conditional
intensity estimation method. A simple illustration of the time
transformation is presented using simulated
data. In Sec.~\ref{sec:OgataTests}, the ``Ogata's tests battery'' is
presented, discussed and illustrated with the simulated data set of
the previous section. In Sec.~\ref{sec:newTest}, a new test, the
``Wiener process test'', is proposed based on Donsker's Theorem. The
Theorem is first stated and a description of how to use it on time
transformed data is given. A ``geometrical'' interpretation is
sketched using the simulated data set. The finite sample properties of
the test are studied next with Monte Carlo simulations. In
Sec.~\ref{sec:realData}, real data are used to
illustrate the performances of the Ogata's tests and of the Wiener
process test. Sec.~\ref{sec:conclusions} summarizes our findings and
discusses them.  

\section{An heuristic approach to the time transformation}
\label{sec:heuristicTimeTransformation}

A simple and elegant proof of the time transformation / time rescaling
theorem appears in~\cite{BrownEtAl_2002}. We are going to present here
an heuristic approach to this key result introducing thereby our
notations. Following~\cite{Brillinger_1988b} we define three quantities:
\begin{itemize}
\item \textbf{Counting Process}: For points $\{ t_j \}$ randomly
  scattered along a line, the counting process $N(t)$ gives the
  number of points observed in the interval $(0,t]$:
  \begin{equation}
    \label{eq:CountingProcessDefinition}
    N(t) = \sharp \{ t_j \; \mathrm{with} \; 0 < t_j \leq t \}
  \end{equation}
  where $\sharp$ stands for the number of elements of
  a set.
\item \textbf{History}: The history, $\mathcal{H}_t$, consists of the
  variates determined up to and including time $t$ that are
  necessary to describe the evolution of the counting
  process. $\mathcal{H}_t$ can include all or part of the neuron's
  discharge up to $t$ but also the discharge sequences of other
  neurons recorded simultaneously, the elapsed time since the onset
  of a stimulus, the nature of the stimulus, etc. One of the major
  problems facing the neuroscientist analysing spike trains is the
  determination of what constitutes $\mathcal{H}_t$ for the data at
  hand. A pre-requisite for practical applications of the approach
  described in this paper is that $\mathcal{H}_t$ involves only a
  finite (but possibly random) time period prior to $t$. In the sequel
  we will by convention set the origin of time at the actual
  experimental time at which $\mathcal{H}_t$ becomes defined. For
  instance if we work with a model requiring the times of the last two
  previous spikes, then the $t=0$ of our counting process definition
  above is the time of the second event / spike. If we consider
  different models with different $\mathcal{H}_t$, the time origin is
  the largest of the ``real times'' giving the origin of each model.
\item \textbf{Conditional Intensity}: For the process $N$ and history
  $\mathcal{H}_t$, the conditional intensity at time $t$ is defined by:
  \begin{equation}
    \label{eq:ciDefinition}
    \lambda (t \mid \mathcal{H}_t) = \lim_{\delta \downarrow 0} \frac{\mathrm{Prob} \{
      N(t,t+\delta) - N(t) = 1 \mid \mathcal{H}_t \}}{\delta}
  \end{equation}
\end{itemize}

As far as we know, most of the presently published attempts to
estimate the \ci~involve a discretization (binning) of the time
axis~\cite{JohnsonSwami_1983,Brillinger_1988b,ChornoboyEtAl_1988,KassVentura_2001,TruccoloEtAl_2005,OkatanEtAl_2005,TruccoloDonoghue_2007}. We
will follow this path here and find the probability density
of the interval between two successive events, $I_j =
t_{j+1}-t_j$. Defining $\delta = \frac{t_{j+1}-t_j}{K}$, where $K \in
\mathbb{N}^{\ast}$. We first write the probability of the interval as the
following product:
\begin{eqnarray}
  \label{eq:Pisi1}
  \mathrm{Pr}\{I_j = t_{j+1} - t_j\} & = & \mathrm{Pr}\{N(t_j+\delta) -
  N(t_j)=0 \mid \mathcal{H}_{t_j}\} \cdot {} \nonumber \\
  & & {} \cdot \mathrm{Pr}\{N(t_j+2\,\delta)-N(t_j+\delta)=0 \mid
  \mathcal{H}_{t_{j+\delta}}\} \cdots {} \nonumber \\
  & & {} \cdots \mathrm{Pr}\{N(t_j+K\,\delta)-N(t_j + (K-1) \,
  \delta)=0 \mid
  \mathcal{H}_{t_{j+ (K-1) \, \delta}}\} \cdot {} \nonumber \\
  & & {} \cdot \mathrm{Pr}\{N(t_j+(K+1)\delta)-N(t_j + K \,
  \delta)=1 \mid \mathcal{H}_{t_{j+ K \delta}}\}
\end{eqnarray}
Then using Eq.~\ref{eq:ciDefinition} we can, for $K$ large enough,
consider only two possible outcomes per bin, there is either no event
or one event. In other words we interpret Eq.~\ref{eq:ciDefinition} as
meaning:
\begin{eqnarray}
  \mathrm{Prob} \{N(t,t+\delta) - N(t) = 0 \mid \mathcal{H}_t \} & = &
  1 - \lambda (t \mid \mathcal{H}_t) \, \delta + \mathrm{o}(\delta)
  \nonumber \\
  \mathrm{Prob} \{N(t,t+\delta) - N(t) = 1 \mid \mathcal{H}_t \} & = &
  \lambda (t \mid \mathcal{H}_t) \, \delta + \mathrm{o}(\delta)
  \nonumber \\
  \mathrm{Prob} \{N(t,t+\delta) - N(t) > 1 \mid \mathcal{H}_t \} & = &
  \mathrm{o}(\delta) \nonumber
\end{eqnarray}
where $\mathrm{o}(\delta)$ is such that $\lim_{\delta \to
  0}\frac{\mathrm{o}(\delta)}{\delta} = 0$. The interval's probability (Eq.~\ref{eq:Pisi1}) becomes therefore the outcome of
a sequence of Bernoulli trials, each with an inhomogeneous success
probability given by $\lambda_i \, \delta + \mathrm{o}(\delta)$,
where, $\lambda_i = \lambda (t_j + i \, \delta \mid \mathcal{H}_{t_j +
  i \, \delta})$ and we get:
\begin{equation}
  \label{eq:Pisi2}
  \mathrm{Pr}\{I_j = t_{j+1} - t_j\} = \big( \prod_{k=1}^{K} (1 -
  \lambda_k \, \delta + \mathrm{o}(\delta)) \big) \, (\lambda_{K+1} \, \delta + \mathrm{o}(\delta))
\end{equation}
We can rewrite the first term on the right hand side as:
\begin{eqnarray}
  \label{eq:Pisi3}
  \prod_{k=1}^{K} (1-\lambda_k \, \delta +
    \mathrm{o}(\delta)) & = & \exp \log \prod_{k=1}^{K} (1-\lambda_k \, \delta +
    \mathrm{o}(\delta)) \nonumber \\
  & = & \exp \sum_{k=1}^{K} \log (1-\lambda_k \, \delta +
    \mathrm{o}(\delta)) \nonumber \\
  & = & \exp \sum_{k=1}^{K} (-\lambda_k \, \delta +
    \mathrm{o}(\delta)) \nonumber \\
  & = & \exp (- \sum_{k=1}^{K} \lambda_k \, \delta) \cdot \exp \big(K \,
  \mathrm{o}(\delta) \big) \nonumber
\end{eqnarray}
Using the continuity of the exponential function, the definition
of the Riemann's integral, the definition of $\delta$ and the property of the $\mathrm{o}()$
function we can take the limit when $K$ goes to $\infty$ on both sides
of Eq.~\ref{eq:Pisi3} to get:
\begin{equation}
  \label{eq:Pisi4}
  \lim_{K \to \infty} \, \prod_{k=1}^{K} (1-\lambda_k \, \delta +
    \mathrm{o}(\delta)) = \exp - \int_{t_j}^{t_{j+1}} \lambda (t \mid
    \mathcal{H}_t) \, dt
\end{equation} 
And the probability density of the interval becomes:
\begin{equation}
  \label{eq:Pisi5}
  \lim_{K \to \infty} \, \frac{\mathrm{Pr}\{I_j = t_{j+1} -
    t_j\}}{\frac{t_{j+1} - t_j}{K}} = \lambda (t_{j+1} \mid
  \mathcal{H}_{t_{j+1}}) \, \exp - \int_{t_j}^{t_{j+1}} \lambda (t \mid
    \mathcal{H}_t) \, dt
\end{equation}
If we now define the \CI~by:
\begin{equation}
  \label{eq:CIDefinition}
  \Lambda(t) = \int_{u=0}^t \lambda (u \mid \mathcal{H}_u) \, du
\end{equation}
We see that $\Lambda$ is increasing since by definition
(Eq.~\ref{eq:ciDefinition}) $\lambda > 0$\footnote{Actual neurons
  exhibit a ``refractory period'', that is, a minimal duration between
  two successive spikes. One could therefore be tempted to allow
  $\lambda = 0$ on a small time interval following a spike. We can
  cope with this potential problem by making $\lambda$ very small but
  non null leading to models which would be indistinguishable from models
  with an absolute refractory period when applied to real world
  (finite) data.}. We see then that the mapping:
\begin{equation}
  \label{eq:CIMap}
  t \in \mathbb{R}^{+ \, \ast} \to \Lambda(t) \in \mathbb{R}^{+ \, \ast}
\end{equation}
is one to one and we can transform our $\{t_1,\ldots,t_n\}$ into
$\{\Lambda_1=\Lambda(t_1),\ldots,\Lambda_n=\Lambda(t_n)\}$. If we now
consider the probability density of the intervals $t_{j+1}-t_j$ and
$\Lambda_{j+1}-\Lambda_j$ we get:
\begin{eqnarray}
  \label{eq:timeTransformation1}
  \mathrm{p}(t_{j+1}-t_j) \, dt_{j+1} & = & \lambda (t_{j+1} \mid
  \mathcal{H}_{t_{j+1}}) \, \exp \big( - \int_{t_j}^{t_{j+1}} \lambda (t \mid
    \mathcal{H}_t) \, dt \big) \, dt_{j+1} \nonumber \\
    & = & \frac{d \Lambda (t_{j+1})}{dt} \, dt_{j+1} \, \exp - \big(
    \Lambda(t_{j+1}) - \Lambda(t_j) \big) \nonumber \\
    & = & d \Lambda_{j+1} \, \exp - \big(
    \Lambda_{j+1} - \Lambda_j \big)
\end{eqnarray}
That is, \emph{the mapped intervals},
$\Lambda_{j+1} - \Lambda_j$ \emph{follow an exponential distribution
  with rate 1}. This is the substance of the \emph{time
  transformation}~of~\cite{Ogata_1988} and of the \emph{time rescaling
theorem}~of~\cite{BrownEtAl_2002}. This is also the justification of
the point process simulation method of~\cite[Sec. 2.3, pp
281-282]{Johnson_1996}. 

\subsection*{An illustration}
\label{sec:timeTransformationIllustration}

\begin{figure}
  \centering
  \includegraphics[width=1.0\textwidth]{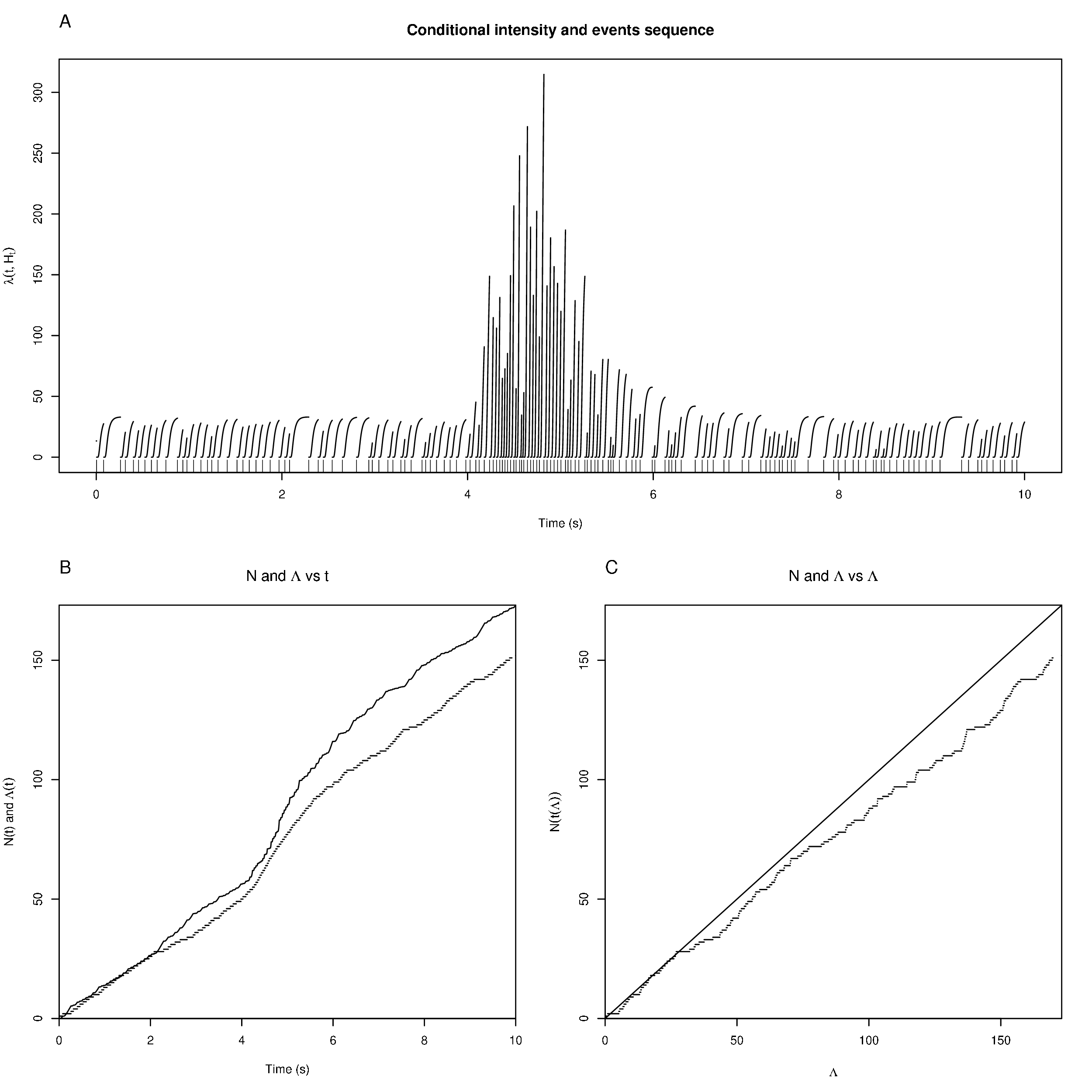}
  \caption{Time transformation illustration. From simulated data with
    the thinning method~\cite{Ogata_1981} according to the \ci~defined
    by Eq.~\ref{eq:ill5}. Parameters used $\mu=0.075$ and $\sigma^2=3$
    in Eq.~\ref{eq:ill1} and~\ref{eq:ill2}; $t_0=4$, $m=5$ and $p=20$
    in Eq.~\ref{eq:ill4}. A, Graph of the \ci~together with the spike
  train represented by a sequence of small vertical bars (one for each
spike at each spike time) at the bottom of the graph. The \ci~is
discontinuous, exhibiting a jump at each spike occurrence. It is left
continuous and converges to 0 on the right-hand side. B, the
\CI, $\Lambda$ (continuous curve) and the counting process, $N$ (step
function) as a function of time. C, Same as B after time transformation.}
  \label{fig:timeTransformationIllustration}
\end{figure}

We illustrate here the time transformation with simulated data. This
is also the occasion to give an example of what an abstract notation
like, $\lambda (t_{j} \mid \mathcal{H}_{t_{j}})$, translates into in
an actual setting. We will therefore consider a counting process whose
\ci~is the product of an inverse-Gaussian hazard function and of the
exponential of a scaled $\chi^2$ density. That would correspond to a
neuron whose spontaneous discharge would be a renewal process with a
inverse-Gaussian \ISI~(\isi) distribution and that would be excited by
``multiplicative'' stimulus (this type of Cox-like model was to our
knowledge first considered in neuroscience by~\cite{JohnsonSwami_1983}). 
More explicitly our \ci~is defined by:
\begin{eqnarray}
  f_{IG}(x) & = & \frac{1}{\sqrt{2\pi x^3 \sigma^2}} \exp \big(-\frac{1}{2}
  \frac{(x-\mu)^2}{x\sigma^2\mu^2}\big) \label{eq:ill1} \\
  F_{IG}(x) & = & \Phi\big( \frac{x-\mu}{\sqrt{x\sigma^2\mu^2}}\big) +
  \exp(\frac{2}{\mu \sigma^2}) \, \Phi\big(
  \frac{-x-\mu}{\sqrt{x\sigma^2\mu^2}}\big) \label{eq:ill2} \\
  h_{IG}(x) & = & \frac{f_{IG}(x)}{1-F_{IG}(x)} \label{eq:ill3} \\
  s(t) & = & p \, f_{\chi^2_5}\big(m \, (t-t_0) \big) \label{eq:ill4} \\
  \lambda (t \mid \mathcal{H}_{t}) & = & h_{IG}(t-t_l) \, \exp
  \big( s(t) \big) \label{eq:ill5}
\end{eqnarray}
where, $\Phi$, is the cumulative distribution function of a standard
normal random variable and $f_{\chi^2_5}$ stands for the probability
density function of a $\chi^2$ random variable with 5 degrees of
freedom and $t_l$ is the occurrence time of the preceding spike. In
this rather simple case the history is limited to: $\mathcal{H}_t= \max
\{t_j : t_j < t\}$. The graph of $\lambda (t \mid \mathcal{H}_{t})$
corresponding to 10 s of simulated data is shown on
Fig.~\ref{fig:timeTransformationIllustration} A, together with the
data. The discontinuous nature of the \ci~appears clearly and is a
general feature of non-Poisson counting
processes. Fig.~\ref{fig:timeTransformationIllustration} B, shows the
graph of the counting process together with the \CI,
$\Lambda(t)$. Fig.~\ref{fig:timeTransformationIllustration} C,
illustrates the time transformation \emph{per se}, the counting
process is now defined on the ``$\Lambda$ scale'' and the integrated
conditional intensity is now a straight line with slope one on this graph.  

\section{The Ogata's tests}
\label{sec:OgataTests}

\begin{figure}
  \centering
  \includegraphics[width=1.0\textwidth]{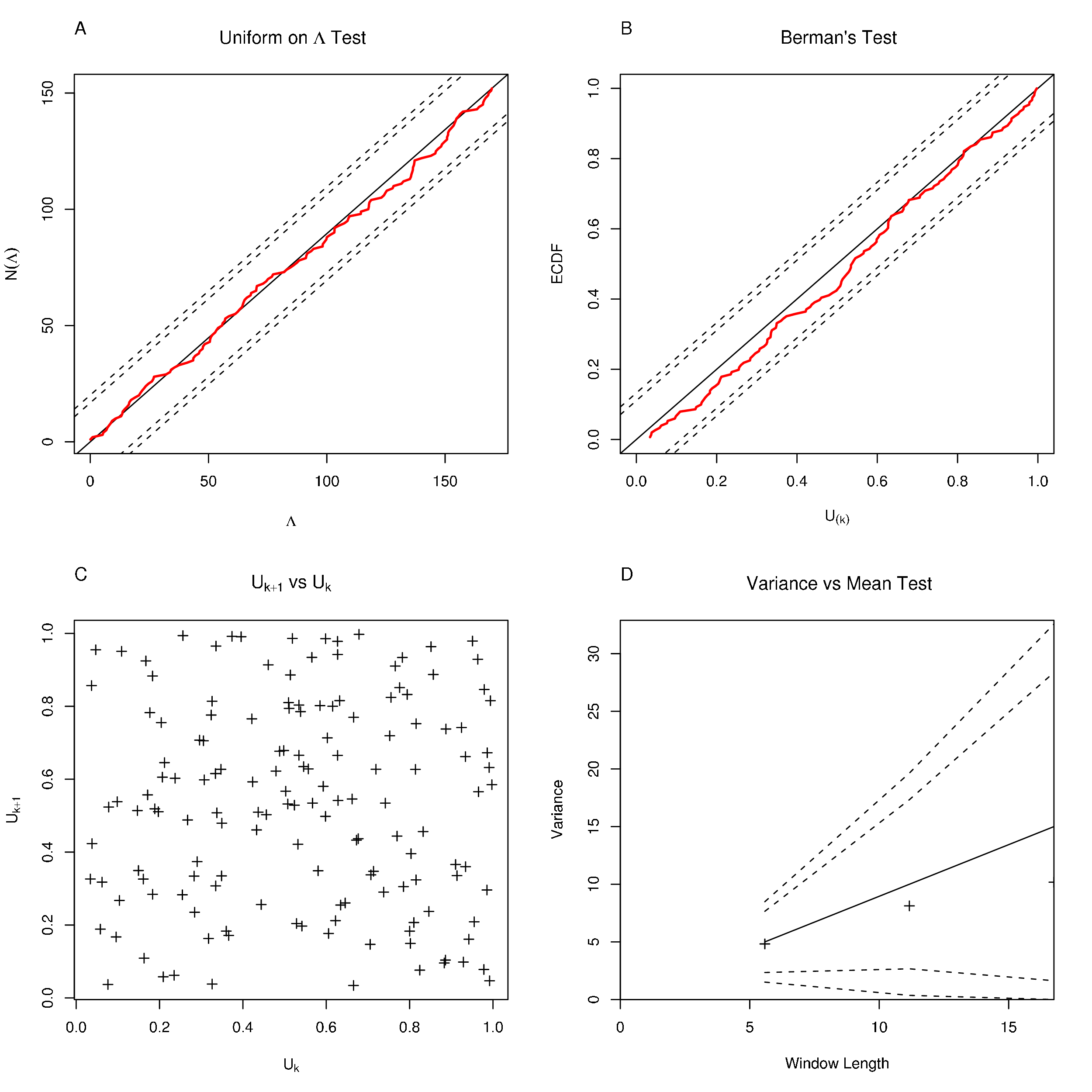}
  \caption{Four of the five Ogata's tests applied to the simulated
    data of Fig.~\ref{fig:timeTransformationIllustration}. }
  \label{fig:OgataTests}
\end{figure}

As soon as we have a model of $\lambda (t \mid \mathcal{H}_{t})$ that
we think (or hope) could be correct we can, following
Ogata~\cite{Ogata_1988}, exploit the time transformation of the
previous section to generate goodness of fit
tests. We start by mapping $\{t_1,\ldots,t_n\}$ onto
$\{\Lambda_1=\Lambda(t_1),\ldots,\Lambda_n=\Lambda(t_n)\}$. Most of
the time this mapping requires a numerical integration of $\lambda (t
\mid \mathcal{H}_{t})$. Then if our model is correct, the mapped
intervals, $\Lambda_{j+1} - \Lambda_{j}$, should be \iid~from an
exponential distribution with rate 1 \emph{and} (equivalently) the
counting process $N(\Lambda)$ should be the realization of a
homogeneous Poisson process with rate 1. In~\cite{Ogata_1988}, Ogata introduced the five following tests:
\begin{enumerate}
\item If a homogeneous Poisson process with rate 1 is observed
  until its $n$th event, then
  the event times, 
  $\{\Lambda_i \}_{i=1}^{n-1}$, have a uniform distribution on
  $(0,\Lambda_n)$~\cite[chap. 2]{CoxLewis_1966}. This uniformity can
  be tested with a Kolmogorov test. This is the first Ogata test~\cite[Fig. 9, p
  19]{Ogata_1988}. It's application to our simulated data is shown on
  Fig.~\ref{fig:OgataTests} A. In the sequel we will refer to this
  test as Ogata's first test or as the ``uniform test''.  
\item The $u_k$ defined, for $k>1$, by:
  \begin{displaymath}
    u_k= 1 - \exp \big(- (\Lambda_k - \Lambda_{k-1}) \big)
  \end{displaymath}
  should be \iid~with a uniform distribution on $(0,1)$. The empirical
  \CDF~of the sorted $\{u_k\}$ can be compared to the 
  \CDF~of the null hypothesis with a Kolmogorov test. This test
  is attributed to 
  Berman in~\cite{Ogata_1988} and is one of the tests proposed and used
  by~\cite{BrownEtAl_2002}. It's application to our simulated data is shown on
  Fig.~\ref{fig:OgataTests} B. This tests implies a sorting of the
  transformed data which would destroy any trace of serial correlation
  in the $u_k$s if one was there. In other words this tests
  that the $u_k$s are identically distributed \emph{not that they are
  independent}. This latter hypothesis has to be checked separately
  which leads us to the third Ogata test.
\item  A plot of $u_{k+1}$ vs $u_{k}$ exhibiting a pattern would be
  inconsistent with the homogeneous Poisson process hypothesis. This
  graphical test is illustrated in
  Fig.~\ref{fig:OgataTests} C for our 
  simulated data. A shortcoming of this test is that it is only
  graphical and that it requires a fair number of events to be
  meaningful. In~\cite{PouzatChaffiol_2008} a nonparametric and more
  quantitative version of this test was proposed.
\item Ogata's fourth test is based on the empirical survivor function
  obtained from the $u_k$s. It's main difference with the second test
  is that only pointwise confidence intervals can be obtained. That
  apart it tests the same thing, namely that the marginal
  distribution of the $u_k$s is exponential with rate 1. Like the
  second test it requires sorting and will fail to show deviations from the
  independence hypothesis. Since this test being redundant with test 2
  it is not shown on Fig.~\ref{fig:OgataTests}.
\item The fifth test is obtained by splitting the transformed time
  axis into $K_w$ non-overlapping
  windows of the same size $w$, counting the number of events in each
  window and getting a mean count $N_w$ and a variance $V_w$ computed
  over the $K_w$ windows. Using a set of increasing window sizes:
  $\{w_1,\ldots,w_L\}$ a graph of $V_w$ as a function of $N_w$ is
  build. If the Poisson process with rate 1 hypothesis is correct the
  result should fall on a straight line going through the
  origin with a unit slope. Pointwise confidence intervals can be
  obtained using the normal approximation of a Poisson distribution as
  shown on Fig.~\ref{fig:OgataTests} C. 
\end{enumerate}

\section{A new test based on Donsker's Theorem: The Wiener Process
  Test} 
\label{sec:newTest}

As we have seen in the previous section each Ogata's test considered
separately has some drawback. This is why Ogota used all of them in
his paper~\cite{Ogata_1988}. It seems therefore reasonable to try to
develop new tests which could replace part of Ogata's tests and / or
behave better than them in some conditions, like when the sample size
is small. We
are going to apply directly Donsker's Theorem to our time
transformed spike trains as an attempt to fulfill this goal.

Following Billingsley~\cite[p 121]{Billingsley_1999}, we start by
defining a functional space:

\paragraph{Definition of $D$}
Let $D=D[0,1]$ be the space of real functions $x$ on $[0,1]$
that are right-continuous and have left-hand limits:
\begin{enumerate}
\item[(i)] For $0 \le t < 1$, $x(t+)=\lim_{s \downarrow t}x(s)$ exists and $x(t+)=x(t)$.
\item[(ii)] For $0 \le t < 1$, $x(t-)=\lim_{s \uparrow t}x(s)$ exists.
\end{enumerate}

\paragraph{Donsker's Theorem}
Given \iid~random variables $\xi_1,\xi_2,\ldots,\xi_n,\ldots$ with mean 0 and
variance $\sigma^2$ defined on a probability space
$(\Omega,\mathcal{F},\mathrm{P})$. We can associate with these random
variables the partial sums $S_n=\xi_1+\cdots +\xi_n$. Let
$X^n(\omega)$ be the function in  $D$ with value:
\begin{equation}
  \label{eq:donsker}
  X^n_t(\omega)=\frac{1}{\sigma \sqrt{n}} S_{\lfloor nt \rfloor}(\omega)
\end{equation}
at $t$, where $\lfloor nt \rfloor$ is the largest integer $\le
nt$. Donsker's Theorem states that: The random functions $X^n$
converge weakly to $W$, the Wiener process on $[0,1]$.

See~\cite[pp 146-147]{Billingsley_1999} for a proof of this theorem. 

\paragraph{Using Donsker's Theorem}

\begin{figure}
  \centering
  \includegraphics[width=1.0\textwidth]{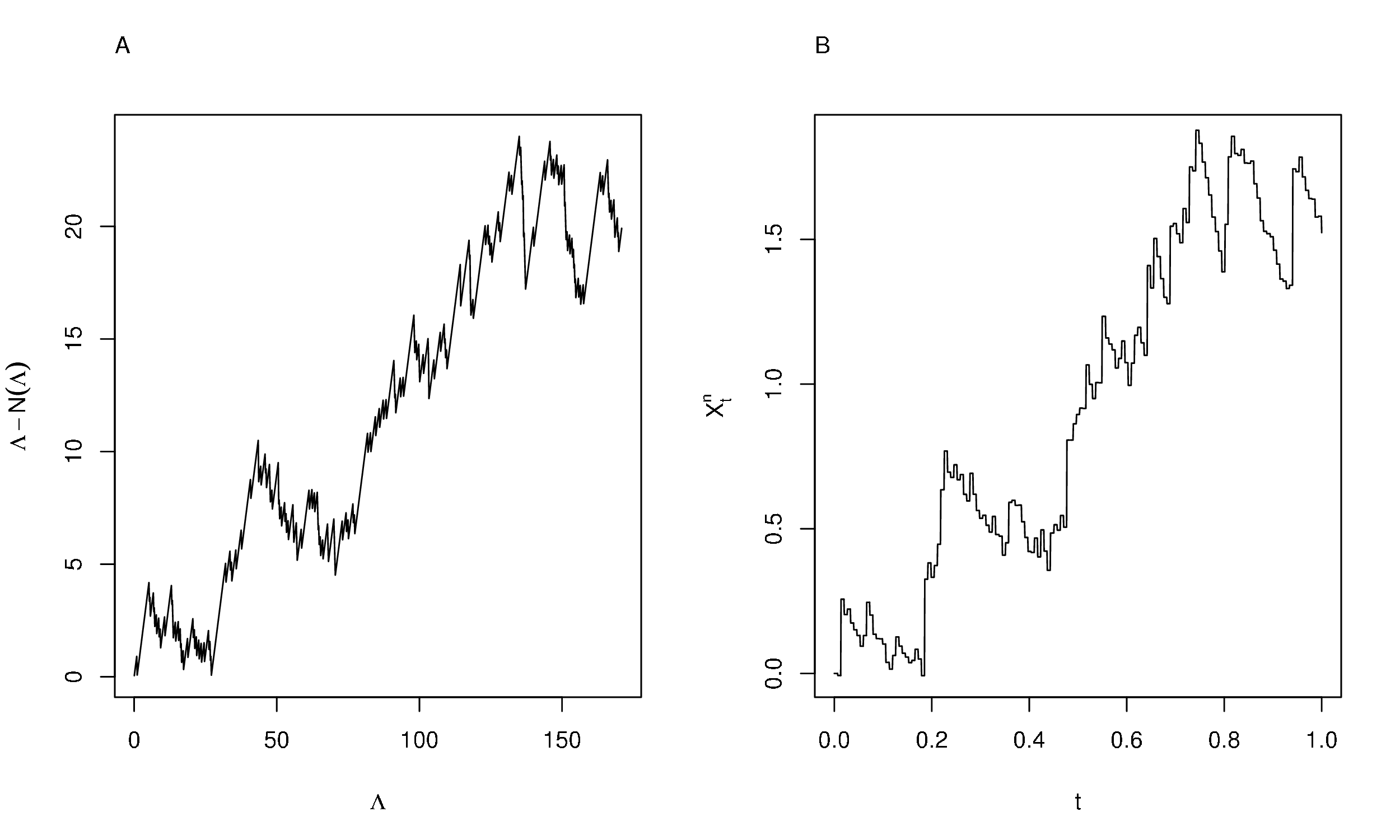}
  \caption{Illustration of the ``mapping'' to a Wiener process. A, The
  counting process has been subtracted from the integrated intensity
  of Fig.~\ref{fig:timeTransformationIllustration} C, on the
  transformed time scale. B, The corresponding realization of the
  $X^n_t$ function of Eq.~\ref{eq:donsker}.}
  \label{fig:wiener1}
\end{figure}

If our model is correct then the $\Lambda_{j+1}-\Lambda_j$ are \iid~random
variables from an exponential distribution with rate 1. Since the mean and
variance of this distribution are both equal to 1, if we define:
\begin{equation}
  \label{eq:meanCorrected}
  \xi_j = \Lambda_{j+1}-\Lambda_j -1
\end{equation}
the $\xi_j$s should be \iid~with mean 0 and variance 1.
Once we have observed the $\Lambda_j$s and therefore the $\xi_j$s we can built the
realization of Eq.~\ref{eq:donsker}. The latter should ``look like'' a
Wiener process on $[0,1]$. Notice that the realization of
Eq.~\ref{eq:donsker} is nearly the same as the function obtained by
subtracting $N(\Lambda)$ from $\Lambda$ on
Fig.~\ref{fig:timeTransformationIllustration} C, before dividing the
abscissa by $n$ and the ordinate by $\sqrt{n}$ as illustrated on
Fig.~\ref{fig:wiener1}. This is also seen by looking at the expression
of the partial sums in this particular case:
\begin{equation}
  \label{eq:partialSums}
  S_n = \Lambda_{n+1}-\Lambda_1 -n
\end{equation}

The next element we need in order to get a proper test is a way to
define a tight region of $[0,1] \times \mathbb{R}$ where, say, 95\% of
the realizations of a Wiener process should be. This turns out to be a
non trivial but luckily solved problem. Kendall et
al~\cite{KendallEtAl_2007} have indeed shown that the boundaries of the
tightest region containing a given fraction of the sample paths of a
Wiener process is are given by a so called Lambert W function. They have moreover
shown that this function can be very well approximated by a square
root function plus an offset: $a + b \, \sqrt{t}$. We next need to
find the $a_{\alpha}$ and $b_{\alpha}$ such that the realizations of a
Wiener process are entirely within the boundaries with a probability
$1-\alpha$. There is no analytical solution to this problem but
efficient numerical solutions are available. Loader and
Deely~\cite{LoaderDeely_1987} have shown that the first passage time
of a Wiener process through an arbitrary boundary $c(t)$ is the
solution of a Volterra integral equation of the first kind. They have
proposed a ``mid-point'' algorithm to find the first passage time
distribution. Their method does moreover provide error bounds when
$c(t) = a + b \, \sqrt{t}$. Then using the ``symmetry'' of the Wiener
process with respect to the time axis we have to find $a_{\alpha}$ and
$b_{\alpha}$ such that the probability to have a first passage $\le 1$
is $\frac{\alpha}{2}$. 
\begin{figure}
  \centering
  \includegraphics[width=0.7\textwidth]{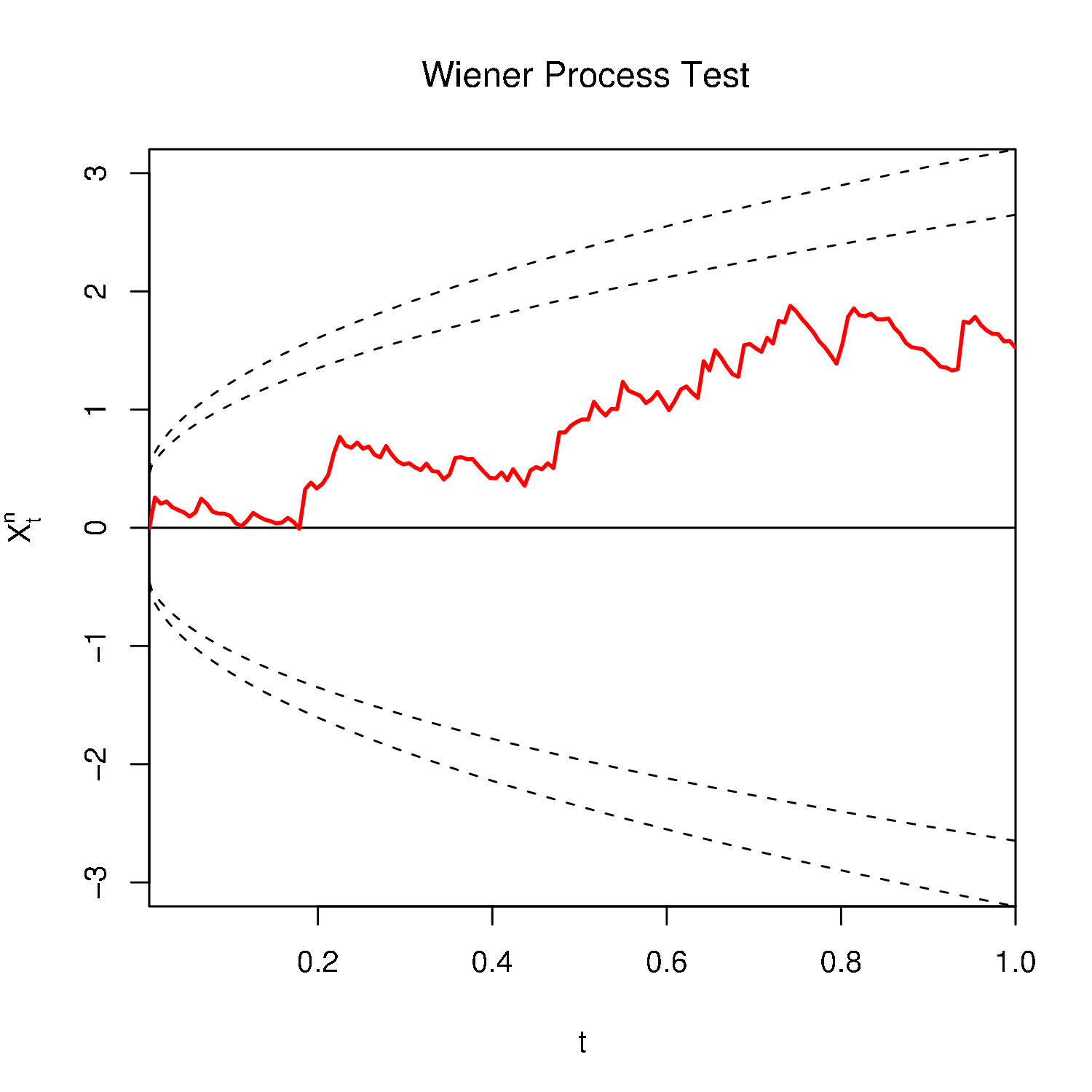}
  \caption{Illustration of the ``Wiener process test''. The same graph
  as Fig.~\ref{fig:wiener1} B is shown with 95\% and 99\% boundaries
  appearing as dotted curves. The boundaries have the form: $a + b \,
  \sqrt{t}$, the coefficients are given in the text.}
  \label{fig:wiener2}
\end{figure}

Using an integration step of 0.001 and,
$a_{0.05}=0.299944595870772$, $b_{0.05}=2.34797018726827$, we get:
$0.9499 < P_{0.95} < 0.9501$. Using
$a_{0.01}=0.313071417065285$, $b_{0.01}=2.88963206734397$, we get:
$0.98998 < P_{0.99} < 0.99002$. In the sequel we will refer to the
comparison of the sample path of $X_t^n$ in Eq.~\ref{eq:donsker} with
boundaries whose general form is: $a_{\alpha}+b_{\alpha} \, \sqrt{t}$
as a \emph{Wiener process test}.

\paragraph{Finite sample properties}

\begin{figure}
  \centering
  \includegraphics[width=1.0\textwidth]{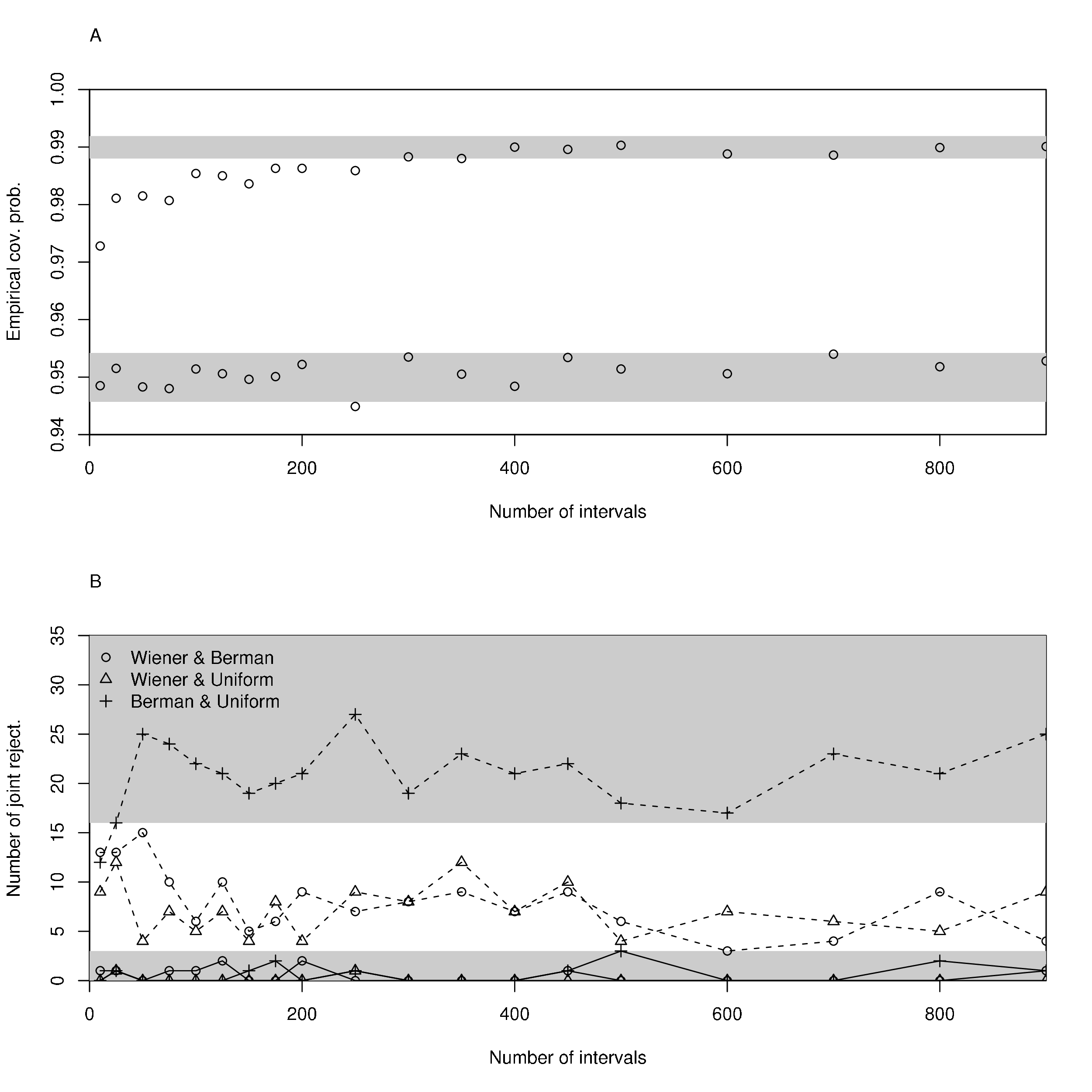}
  \caption{A, Empirical coverage probabilities of the 95 and 99\%
    confidence regions for the Wiener test. Each dot on the graph was
    obtained from 10000 simulated ``experiments''. Each experiment was
    made of a number, $n$, of draws given on the abscissa. Each draw
    was the realization of an exponential distribution with
    rate 1. The procedure described by applying
    Eq.~\ref{eq:meanCorrected} before Eq.~\ref{eq:donsker} was then
    applied. The resulting step function was compared, at the steps
    locations with the boundaries values. The grey band correspond to
    95\% confidence bands for a binomial distribution with 10000 draws
    and a success  probability of 0.95 (bottom) and 0.99 (top). B,
    Number of joint rejections of paired tests. Using the same
    simulations as in A, the number of joint rejection of the ``Wiener
  process test'' and of the ``Berman's test'' (circles), of the ``Wiener
  process test'' and of the ``Uniform test'' (\emph{i.e.}, Ogata's first
  test, triangles), of the ``Berman's test'' and of the ``Uniform
  test'' (crosses) at the 95\% level (dotted lines) and at the 99\%
  level (continuous line). Gray areas, 95\% confidence region for a
  binomial distribution with 10000 trials and a rejection probability
  of $0.05^2$ (top) and $0.01^2$ (bottom). The p values of the
  Kolmogorov statistic of the ``Berman'' and ``Uniform'' tests were
  obtained with the exact method of~\cite{MarsagliaEtAl_2003} and
  implemented in function \texttt{ks.test} of \R.}
  \label{fig:wiener3}
\end{figure}

We now estimate the empirical coverage probability of our two regions
for a range of ``realistic'' sample sizes, from 10 to 900, using Monte
Carlo simulations. The results from 10000 simulated ``experiments''
are shown on Fig.~\ref{fig:wiener3} A. The empirical coverage
probability of the region with a nominal 95\% confidence is at the
nominal level across the whole range studied. The empirical coverage
probability of the region with a nominal 99\% confidence does slightly
worse. The empirical coverage is roughly 98\% for a sample size
smaller than 100, 98.5\% between 100 and 300 and reaches the nominal
value for larger sample sizes. The overall performances of the test
are surprisingly good given the simplicity of its implementation.

Since we are also interested in carrying out multiple testing
using for instance Ogata's first test, Berman's test and our
``Wiener process test'', we studied the dependence of false negatives
under the three possible pairs formed from these three tests. The
results are shown on Fig.~\ref{fig:wiener3} B. Here we see that
Berman's and Ogata's first tests are independent (\emph{i.e.}, the
empirical results are compatible with the hypothesis of independence). But
neither of these tests is independent of the ``Wiener process test''
when a 95\% confidence level (on each individual test) is used. In
both cases (Berman and Ogata's first test) when a trial is rejected at
the 95\% level \emph{it is less likely}~to be also rejected by the
``Wiener process test''. If we combine the three tests at the 99\%
level and decide to reject the null hypothesis if any one of them is
rejected, we obtain an empirical coverage probability of $\sim 0.96$
for a sample size between 10 and 100. Above this size the probability
becomes $\sim 0.97$.  

\paragraph{A remark}

It is clear from the hypothesis of Donsker's theorem that the ``Wiener
process test'' cannot replace Ogata's second test (Berman's test). As long as the
distribution of the mapped intervals has the same first and second
moments than an exponential distribution with rate 1, the data will
pass the test.

\section{Real data examples}
\label{sec:realData}

We illustrate in this section the performances of the tests on real
data. The data, like the software implementing the test and the whole
analysis presented in this article are part of our \STAR~(Spike
Train Analysis with R) package for the
\R\footnote{\url{http://www.r-project.org}}~software. The data were
recorded extracellularly \emph{in vivo}~from the first olfactory relay
of an insect, the cockroach, \emph{Periplaneta americana}. Details
about these recordings and associated analysis methods can be found
in~\cite{Chaffiol_2007,PouzatChaffiol_2008}. The names of the data
sets used in this section are the names used for the corresponding
data in \STAR. The reader can therefore easily reproduce the analysis
presented here. We do moreover provide with \STAR~a
meta-file~\cite{RossiniLeisch_2003,GentlemanTempleLang_2004}, called 
a \emph{vignette}~in \R~terminology, allowing the reader to re-run
exactly the simulation / data analysis of this article.

\subsection{Spontaneous activity data}
\label{sec:spontaneousActivityData}

\begin{figure}
  \centering
  \includegraphics[width=1.0\textwidth]{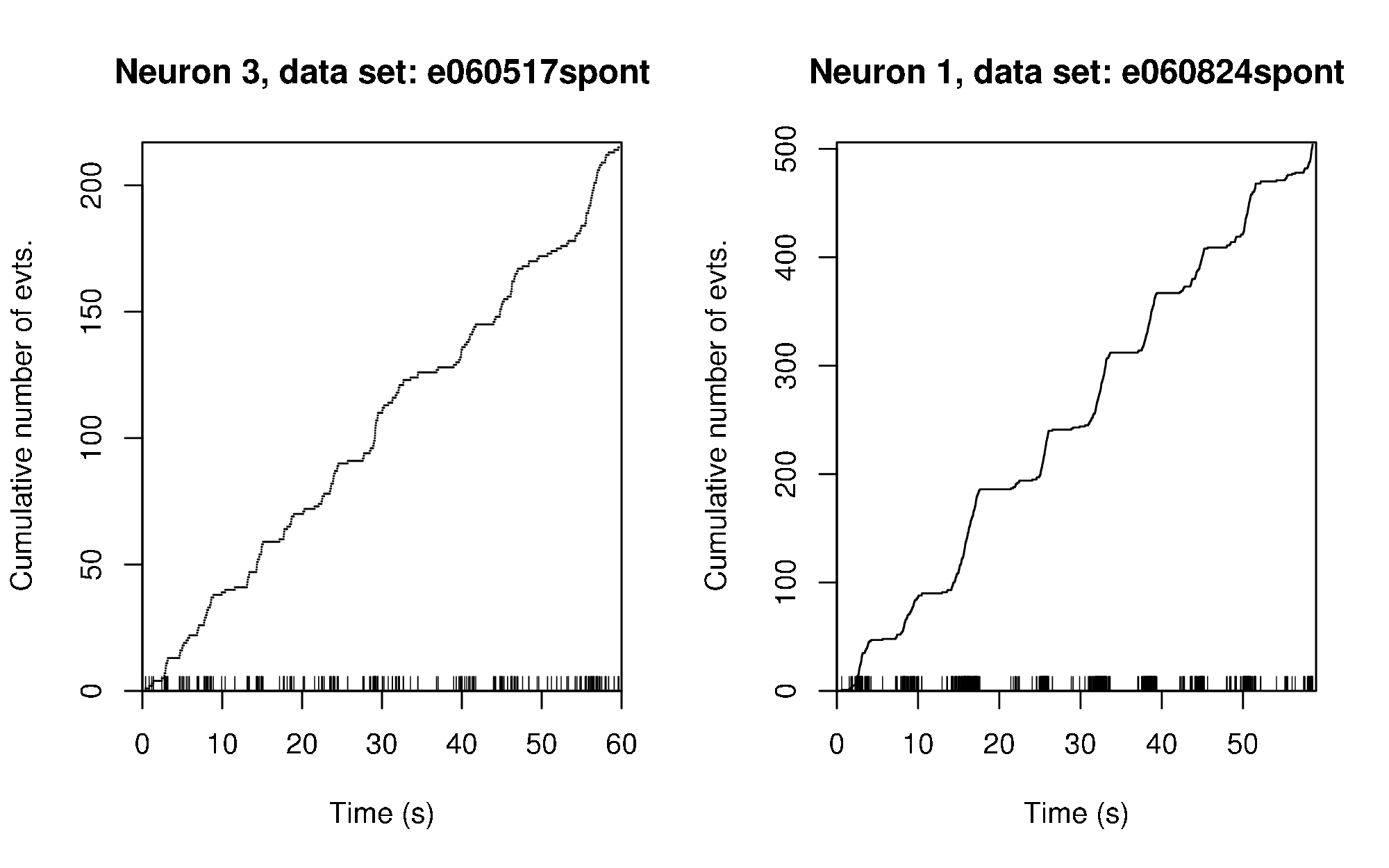}
  \caption{Two actual spike trains. Each plot shows the counting
    process sample path ($N(t)$) associated with a train. The spikes density
    and their number are too high for the discontinuities of $N(t)$ to
  systematically appear. A "raster plot" is drawn at the bottom of each plot. A raster
plot is a spike train representation where the occurrence time of each
spike is marked by a tick (see~\cite{PouzatChaffiol_2008} for details).}
  \label{fig:realdata}
\end{figure}

\begin{figure}
  \centering
  \includegraphics[width=1.0\textwidth]{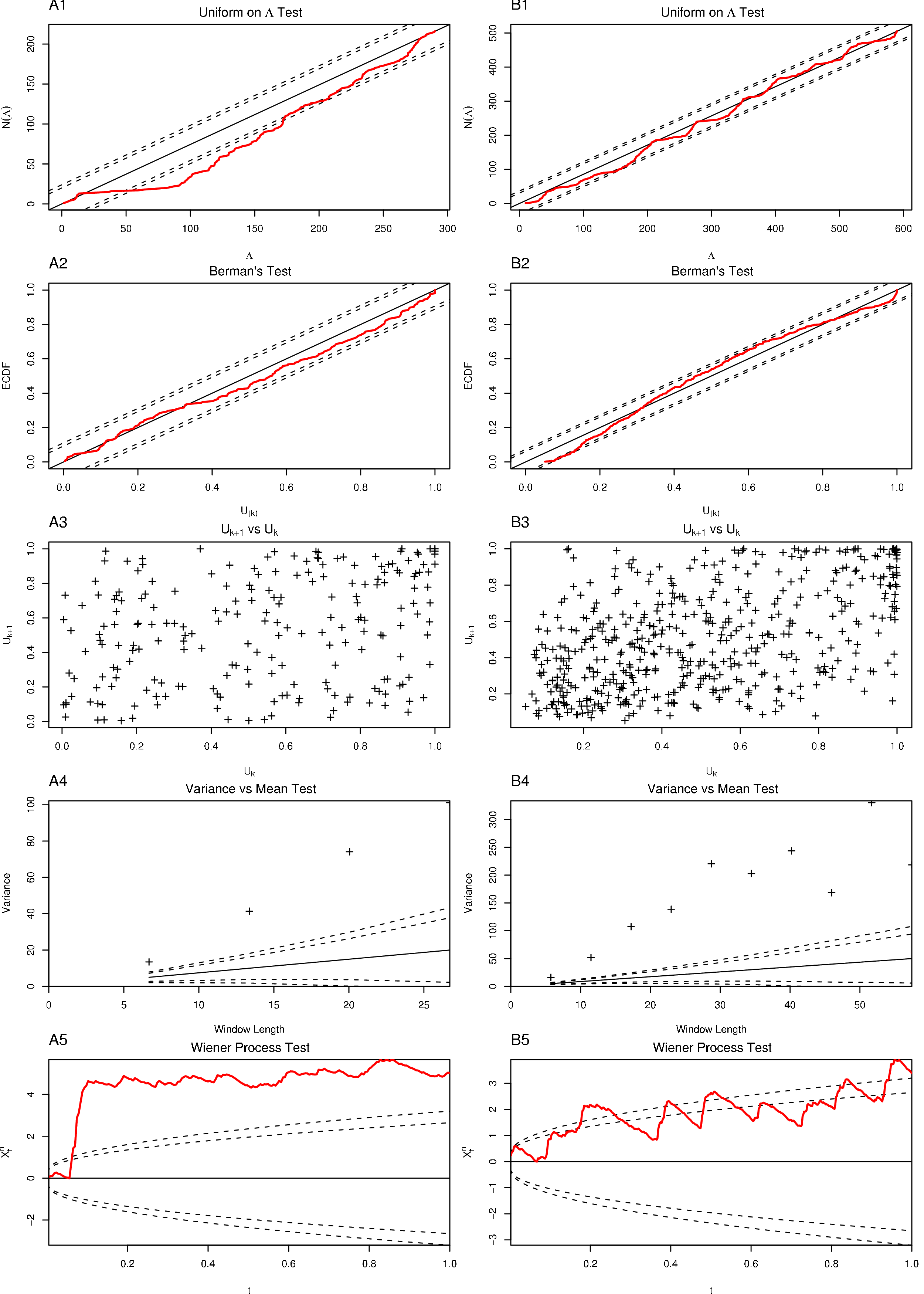}
  \caption{The tests applied after time transformation. A, neuron 3
    from data set \texttt{e060517spont}~fitted with an inverse-Gaussian
    renewal model. B, neuron 1 from data set
    \texttt{e060824spont}~fitted with a log logistic renewal
    model. Notice that the tests are here applied directly to the data
  used to fit the model. The domains defined by the boundaries (dotted
  lines) have therefore not exactly their nominal coverage value.}
  \label{fig:testrealdata}
\end{figure}

Two examples of spike trains recorded during a period of 60 s of
spontaneous activity are shown on Fig.~\ref{fig:realdata}. The spike
trains appear as both a counting process sample path and as a ``raster
plot'' on this figure. A renewal model was fitted with the maximum
likelihood method to each train. The
data from neuron 3 of data set \texttt{e060517spont} were fitted with
an inverse-Gaussian model, while the ones of neuron 1 of data set
\texttt{e060824spont} were fitted with a log logistic model. The spike
times where transformed (Eq.~\ref{eq:CIDefinition}) before applying
Ogata's test 1, 2, 3, 5 and the new Wiener test. The results are shown
on Fig.~\ref{fig:testrealdata}. 

The time transformed spike train of neuron 3 of data set
\texttt{e060517spont} passes Berman's test
(Fig.~\ref{fig:testrealdata} A2) as well as Ogata's third test
(Fig.~\ref{fig:testrealdata} A3) but fails to pass Ogata's tests 1 and
5 (Fig.~\ref{fig:testrealdata} A1 and A4) as well as the new Wiener
process test (Fig.~\ref{fig:testrealdata} A5). The time transformed
spike train of neuron 1 of data set \texttt{e060824spont} passes
Ogata's first test (Fig.~\ref{fig:testrealdata} B1) as well as
Berman's test (Fig.~\ref{fig:testrealdata} B2), but fails Ogata's
tests 3 (Fig.~\ref{fig:testrealdata} B3) and 5
(Fig.~\ref{fig:testrealdata} B4) as well as the new Wiener process
test (Fig.~\ref{fig:testrealdata} B5).

\subsection{Stimulus evoked data}
\label{sec:stimulusEvokedData}

\begin{figure}
  \centering
  \includegraphics[width=0.7\textwidth]{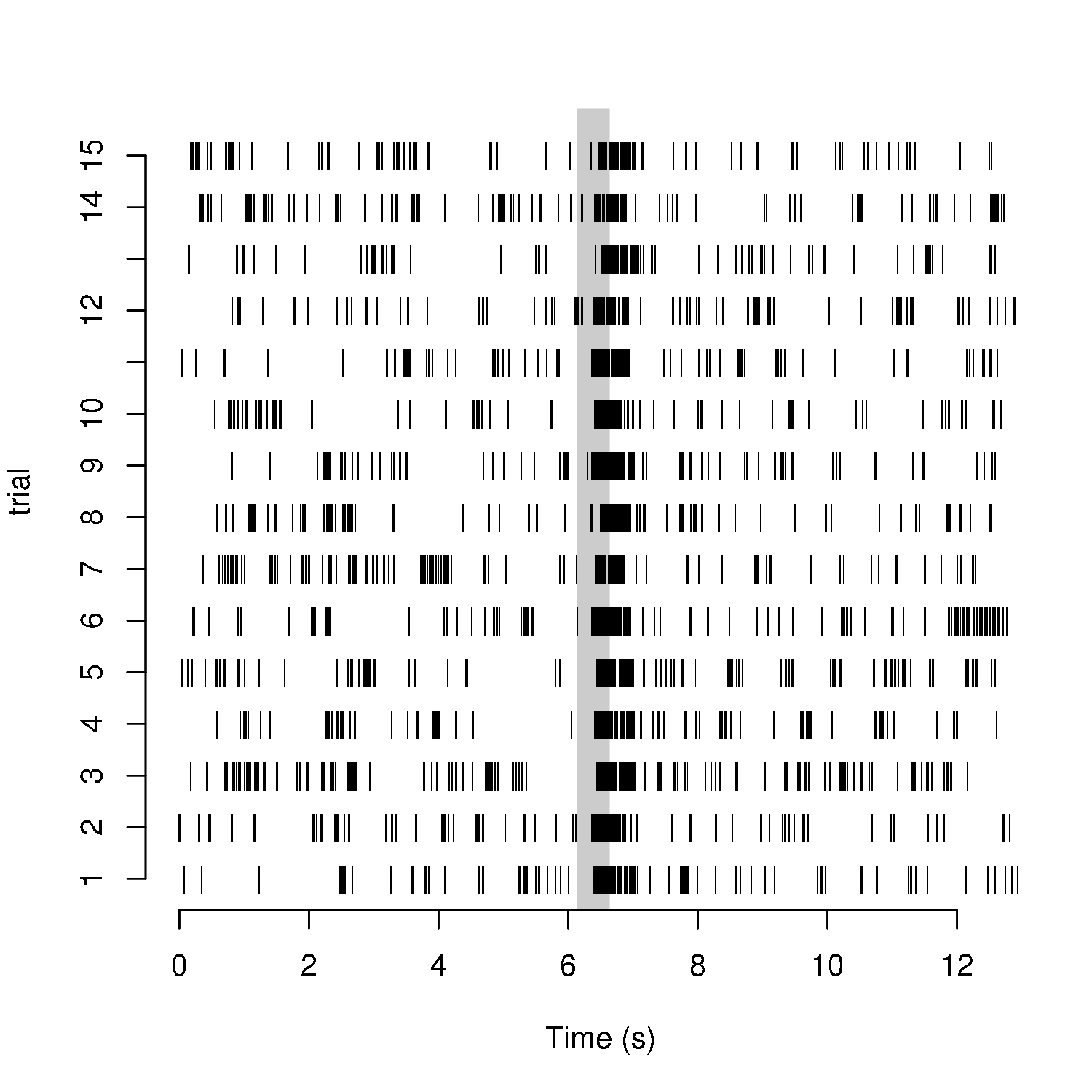}
  \caption{An example of odor evoked responses. Neuron 1 of data set
    \texttt{e070528citronellal} is used. 15 citronellal odor puffs
    were applied successively. Each puff was 0.5
    long~\cite{Chaffiol_2007,PouzatChaffiol_2008} (the gray area
    corresponds to the opening time of the valve delivering the
    odor). Odor presentations were performed 1 mn apart. The spike
    trains of each trial are represented as raster plots. The first
    trial is at the bottom of the graph, the 15th on top.}
  \label{fig:evokeddata}
\end{figure}

\begin{figure}
  \centering
  \includegraphics[width=0.8\textwidth]{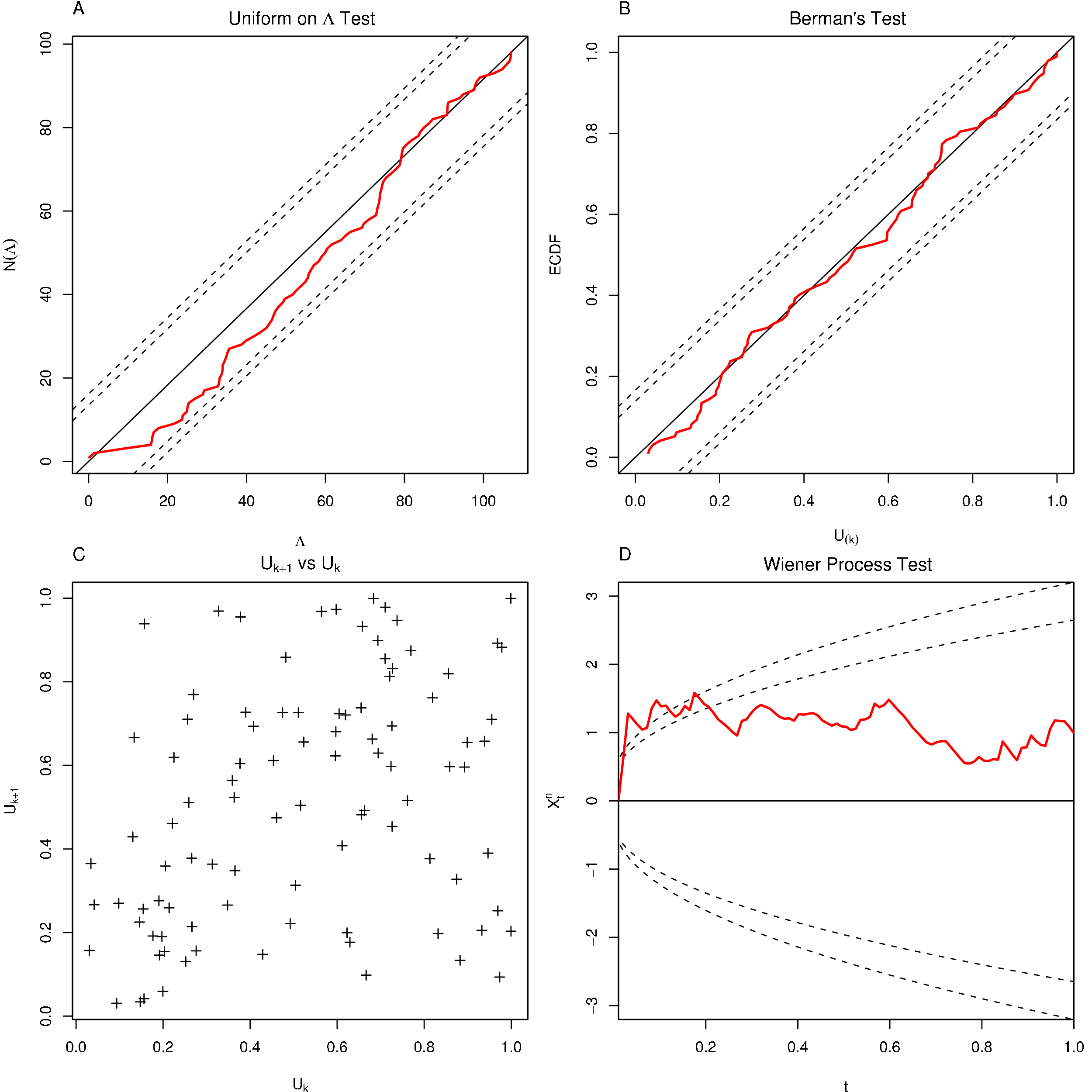}
  \caption{The tests applied after time transformation. An
    inhomogeneous Poisson model was fitted to trial 2 to 15 and used
    to transform the spike times of trial 1.}
  \label{fig:testevokeddata}
\end{figure}

The odor evoked responses of neuron 1 of data set
\texttt{e070528citronellal} are illustrated here. This is the occasion to
use a more sophisticated intensity model, albeit still a wrong one (see
bellow). The ``raw data'' are shown on Fig.~\ref{fig:evokeddata}. An
inhomogeneous Poisson model was fitted to trial 2 to 15 using a
smoothing spline approach~\cite{Gu_2002,PouzatChaffiol_2008}. This
estimated time dependent intensity was then used to transform the spike
times of the first trial. The tests are shown on
Fig.~\ref{fig:testevokeddata}. In this case the train contains 98
spikes making Ogata's test 5 barely applicable (not shown on
figure). Ogata's test 3 (Fig.~\ref{fig:testevokeddata} C) is hard to
interpret since with few spikes and a sparse filling of the graph, the
presence or absence of a pattern is not obvious. Ogata's test 1
(Fig.~\ref{fig:testevokeddata} A) and Berman's test
(Fig.~\ref{fig:testevokeddata} B) are passed but the Wiener process
test (Fig.~\ref{fig:testevokeddata} D) is not.      

\section{Conclusions}
\label{sec:conclusions}

Ogata~\cite{Ogata_1988} introduced a tests battery for counting
process models. He made moreover rather clear through his use of the tests
that more than a single one of them should be used in practice. Brown
et al~\cite{BrownEtAl_2002} introduced in the neuroscience literature
Berman's test (Ogata's second test) which has become the only test used
(when any is used) in this field. We could also remark that only
one case of a model fitted to real data actually passing this
test has ever been published~\cite[Fig. 1A]{BrownEtAl_2002}. Clearly
our examples of Fig.~\ref{fig:testrealdata} A2 and B2 and of
Fig.~\ref{fig:testevokeddata} B, would multiply this number by
four. But we have to stick to Ogata's implicit message. Our choice of
examples reiterates this point showing cases where Berman's test is
passed. We do not want to imply that Berman's test is a ``bad'' test
but to warn the neuroscience community that misleading conclusions can
be obtained when \emph{only}~this test is used.

We have also introduced the ``Wiener process test'' a direct
application of Donsker's Theorem to the time transformed spike
trains. The test is simple to implement (Eq.~\ref{eq:meanCorrected}
and~\ref{eq:donsker}) and exhibits good finite sample properties
(Fig.~\ref{fig:wiener3}). It is clearly not a replacement for the
Ogata's tests battery but a complement (Fig.~\ref{fig:testrealdata}
and~\ref{fig:testevokeddata}). It's finite sample properties make it
particularly useful in situations were a small number of spikes makes
the interpretation of Ogata's tests 3 and 5 potentially ambiguous
(Fig.~\ref{fig:testevokeddata}). 

\section*{Acknowledgments}
\label{sec:acknowledgments}

We thank Vilmos Prokaj for pointing Donsker's Theorem to us as well as
for giving us the reference to Billingsley's book. We thank Olivier
Faugeras and Jonathan Touboul also for mentioning Donsker's Theorem
and Chong Gu for
comments on the manuscript. C. Pouzat was partly supported by a grant
from the Decrypton project, a partnership between the Association Fran\c caise contre les
Myopathies (AFM), IBM and the Centre National de la Recherche
Scientifique (CNRS) and by a CNRS GDR grant (Syst\`eme Multi-\'el\'ectrodes
et traitement du signal appliqu\'es \`a l'\'etude des r\'eseaux de
Neurones). A. Chaffiol was supported by a 
fellowship from the Commissariat \`a l'\'Energie Atomique (CEA).

\pagebreak
\bibliographystyle{plain}
\bibliography{OnTests}

\begin{thebibliography}{10}

\bibitem{Billingsley_1999}
Patrick Billingsley.
\newblock {\em Convergence of Probability Measures}.
\newblock Wiley - Interscience, 1999.

\bibitem{Brillinger_1988b}
D.~R. Brillinger.
\newblock Maximum likelihood analysis of spike trains of interacting nerve
  cells.
\newblock {\em Biol Cybern}, 59(3):189--200, 1988.

\bibitem{BrownEtAl_2002}
Emery~N Brown, Riccardo Barbieri, Val\'{e}rie Ventura, Robert~E Kass, and
  Loren~M Frank.
\newblock The time-rescaling theorem and its application to neural spike train
  data analysis.
\newblock {\em Neural Comput}, 14(2):325--346, Feb 2002.
\newblock Available from:
  \url{http://www.stat.cmu.edu/~kass/papers/rescaling.pdf}.

\bibitem{Chaffiol_2007}
Antoine Chaffiol.
\newblock {\em \'Etude exp\'erimentale de la repr\'esentation des odeurs dans
  le lobe antennaire de \emph{Periplaneta americana}.}
\newblock PhD thesis, Universit\'e Paris XIII, 2007.
\newblock Available at:
  \url{http://www.biomedicale.univ-paris5.fr/physcerv/C_Pouzat/STAR_folder/The%
seAntoineChaffiol.pdf}.

\bibitem{ChornoboyEtAl_1988}
E.~S. Chornoboy, L.~P. Schramm, and A.~F. Karr.
\newblock Maximum likelihood identification of neural point process systems.
\newblock {\em Biol Cybern}, 59(4-5):265--275, 1988.

\bibitem{CoxLewis_1966}
D.~R. Cox and P.~A.~W. Lewis.
\newblock {\em The Statistical Analysis of Series of Events}.
\newblock John Wiley \& Sons, 1966.

\bibitem{GentlemanTempleLang_2004}
Robert Gentleman and Duncan Temple~Lang.
\newblock Statistical {A}nalyses and {R}eproducible {R}esearch.
\newblock Working Paper~2, Bioconductor Project Working Papers, 29 May 2004.
\newblock Available at: \url{http://www.bepress.com/bioconductor/paper2/}.

\bibitem{Gu_2002}
Chong Gu.
\newblock {\em Smoothing Spline Anova Models}.
\newblock Springer, 2002.

\bibitem{Johnson_1996}
D.H. Johnson.
\newblock Point process models of single-neuron discharges.
\newblock {\em J. Computational Neuroscience}, 3(4):275--299, 1996.

\bibitem{JohnsonSwami_1983}
Don~H. Johnson and Ananthram Swami.
\newblock The transmission of signals by auditory-nerve fiber discharge
  patterns.
\newblock {\em J. Acoust. Soc. Am.}, 74(2):493--501, August 1983.

\bibitem{KassVentura_2001}
R.~E. Kass and V.~Ventura.
\newblock A spike-train probability model.
\newblock {\em Neural Comput.}, 13(8):1713--1720, 2001.
\newblock Available from: \url{http://www.stat.cmu.edu/~kass/papers/imi.pdf}.

\bibitem{KendallEtAl_2007}
W.S. Kendall, J.M. Marin, and C.P. Robert.
\newblock Brownian {C}onfidence {B}ands on {M}onte {C}arlo {O}utput.
\newblock {\em Statistics and Computing}, 17(1):1--10, 2007.
\newblock Preprint available at:
  \url{http://www.ceremade.dauphine.fr/\%7Exian/kmr04.rev.pdf}.

\bibitem{LoaderDeely_1987}
C.~R. Loader and J.~J. Deely.
\newblock Computations of boundary crossing probabilities for the {W}iener
  process.
\newblock {\em J. Statist. Comput. Simulation}, 27(2):95--105, 1987.

\bibitem{MarsagliaEtAl_2003}
George Marsaglia, Wai~Wan Tsang, and Jingbo Wang.
\newblock Evaluating {K}olmogorov's {D}istribution.
\newblock {\em Journal of Statistical Software}, 8(18):1--4, 11 2003.

\bibitem{Ogata_1981}
Y.~Ogata.
\newblock On lewis' simulation method for point processes.
\newblock {\em IEEE Transactions on Information Theory}, IT-29:23--31, 1981.

\bibitem{Ogata_1988}
Yosihiko Ogata.
\newblock Statistical {M}odels for {E}arthquake {O}ccurrences and {R}esidual
  {A}nalysis for {P}oint {P}rocesses.
\newblock {\em Journal of the American Statistical Association}, 83(401):9--27,
  1988.

\bibitem{OkatanEtAl_2005}
Murat Okatan, Matthew~A. Wilson, and Emery~N. Brown.
\newblock Analyzing {F}unctional {C}onnectivity {U}sing a {N}etwork
  {L}ikelihood {M}odel of {E}nsemble {N}eural {S}piking {A}ctivity.
\newblock {\em Neural Computation}, 17(9):1927--1961, September 2005.

\bibitem{PerkelEtAl_1967}
D.~H. Perkel, G.~L. Gerstein, and G.~P. Moore.
\newblock Neuronal spike trains and stochastic point processes. {I} the single
  spike train.
\newblock {\em Biophys. J.}, 7:391--418, 1967.
\newblock Available from:
  \url{http://www.pubmedcentral.nih.gov/articlerender.fcgi?tool=pubmed&pubmedi%
d=4292791}.

\bibitem{PerkelEtAl_1967b}
D.~H. Perkel, G.~L. Gerstein, and G.~P. Moore.
\newblock Neuronal spike trains and stochastic point processes. {II}
  simultaneous spike trains.
\newblock {\em Biophys. J.}, 7:419--440, 1967.
\newblock Available from:
  \url{http://www.pubmedcentral.nih.gov/articlerender.fcgi?tool=pubmed&pubmedi%
d=4292792}.

\bibitem{PouzatChaffiol_2008}
Christophe Pouzat and Antoine Chaffiol.
\newblock Automatic spike train analysis and report generation. an
  implementation with {R}, {R2HTML} and {STAR}.
\newblock Submitted manuscript. Pre-print distributed with the STAR package:
  \url{http://cran.at.r-project.org/web/packages/STAR/index.html}, 2008.

\bibitem{RossiniLeisch_2003}
Anthony Rossini and Friedrich Leisch.
\newblock Literate {S}tatistical {P}ractice.
\newblock UW Biostatistics Working Paper Series 194, University of Washington,
  2003.

\bibitem{TruccoloDonoghue_2007}
Wilson Truccolo and John~P. Donoghue.
\newblock Nonparametric modeling of neural point processes via stochastic
  gradient boosting regression.
\newblock {\em Neural Computation}, 19(3):672--705, 2007.
\newblock Available at:
  \url{http://donoghue.neuro.brown.edu/pubs/Truccolo_and_Donoghue_Neural_Comp_%
2007.pdf}.

\bibitem{TruccoloEtAl_2005}
Wilson Truccolo, Uri~T. Eden, Matthew~R. Fellows, John~P. Donoghue, and
  Emery~N. Brown.
\newblock A {P}oint {P}rocess {F}ramework for {R}elating {N}eural {S}piking
  {A}ctivity to {S}piking {H}istory, {N}eural {E}nsemble and {E}xtrinsic
  {C}ovariate {E}ffects.
\newblock {\em J Neurophysiol}, 93:1074--1089, February 2005.
\newblock Available at:
  \url{http://jn.physiology.org/cgi/content/full/93/2/1074}.

\end{thebibliography}

\pagebreak



\end{document}